\def\be{-E/A}
\def\bge{\begin{equation}}
\def\ene{\end{equation}}
\def\bg{\begin{eqnarray}}
\def\en{\end{eqnarray}}

\documentclass[prc,preprint,showpacs,nofootinbib]{revtex4}
\usepackage{graphicx}
\begin{document}

\title{Binding energies and modelling of nuclei\\
in semiclassical simulations 
}

\author{M. \'{A}ngeles P\'{e}rez-Garc\'{i}a~\footnote{mperezga@usal.es}, 
K. Tsushima~\footnote{tsushima@usal.es}, 
A. Valcarce~\footnote{valcarce@usal.es}}

\affiliation{Departamento de F\'{i}sica Fundamental and 
Instituto Universitario de
F\'{i}sica Fundamental y Matem\'{a}ticas, IUFFyM, \\Universidad de Salamanca, 
Plaza de la Merced s/n 37008 Salamanca}

\date{\today}
% Deleting this command produces today's date.

\begin{abstract}
We study the binding energies of spin-isospin saturated nuclei with 
nucleon number $8 \le A \le 100$ in semiclassical Monte Carlo many-body 
simulations.
The model Hamiltonian consists of, 
(i) nucleon kinetic energy, 
(ii) a nucleon-nucleon interaction potential,  
and (iii) an effective Pauli potential which depends on density.
The basic ingredients of the nucleon-nucleon potential are, 
a short-range repulsion, and a medium-range attraction.
Our results demonstrate that one can always expect 
to obtain the empirical binding energies for a set of 
nuclei by introducing a proper density dependent Pauli 
potential in terms of a single variable, the nucleon number, A.    
The present work shows that in the suggested procedure there is a delicate counterbalance of kinetic and potential energetic contributions allowing a good reproduction of the experimental nuclear binding energies. This type of calculations may be of interest in further reproduction of other properties of nuclei such us radii and also exotic nuclei.
\end{abstract}
\vspace{1pc}
\pacs{07.05.Tp,21.10.D,21.65.+f}
%07.05.Tp: Computer modeling and simulation
%21.10.D: Nuclear binding energy
%21.65.+f: Nuclear matter
%\keywords{Monte Carlo simulation, Nuclear binding energy, 
%Nuclear many-body system, Pauli potential}

\maketitle

In our exploratory work~\cite{aka}, we studied the nuclear 
binding energies for medium mass nuclei 
with nucleon number $8 \le A \le 44$ in semiclassical simulations 
via Monte Carlo many-body techniques. 
The purpose of that work was to study the role of an effective Pauli 
potential, which is often adopted in semiclassical 
simulations of many-nucleon systems. 
It was demonstrated that the empirical binding energies 
for these nuclei can be reproduced satisfactorily  
using the Pauli potential, where the density dependence 
is parameterized by one variable, the Fermi momentum. 
The agreement with the empirical binding energies  
was excellent in spite of the simplicity of the model.
The conclusion of our previous work is rather general, 
it does not depend on the detail of the NN potential  
provided a short-range repulsion and medium-range attraction 
are included.
One can always expect to find a proper counter balancing 
density dependent Pauli potential to reproduce 
the empirical binding energies. 
Although the model is not based on the fundamental 
physics of strong interaction QCD, it gives a possible guidance for
treating complicated many-nucleon systems in a simple, 
practical manner in semiclassical simulations.
This may be very helpful to study the many-nucleon systems such as   
the {\it pasta phase}~\cite{pasta,Watanabe,Maruyama,hor} and 
neutron halo nuclei~\cite{Tanihata}.

In this study, we are able to extend our previous work~\cite{aka} to 
treat a wider range of nuclei with $8 \le A \le 100$ 
for spin-isospin saturated $Z=N$ (even Z and N) nuclei. We show that the density dependence of the Pauli potential 
can be well parameterized in the whole range in terms of one single variable, the nucleon number $A$. 
The Fermi momentum is no longer a good parameter due to the fact that the average Fermi momenta in nuclei 
with $A > 50$ saturate to a value $\simeq 260$ MeV/c 
(e.g., $265$ MeV/c for a $^{208}$Pb nucleus)~\cite{moniz}.
The present result generalizes our previous conclusion  
that the density dependence of the Pauli potential is crucial to 
reproduce the empirical nuclear binding energies.
 It is important to explore heavier systems since stability of this type of procedures must be tested with respect to increasing number of nucleons. We have shown in Ref.~\cite{aka} that in the limit of infinite symmetric nuclar matter in these many-body simulations one should obtain the corresponding binding energy around $-16$ MeV. Also in other simulations as in heavy-ion or pasta phases one usually has a wide mass distribution of clusters.
In addition, we explicitly show that it is possible to reproduce the 
empirical binding energies using different NN interaction potentials. 
Thus, it suggests a simple and  pragmatic 
procedure in modelling a set of nuclei calibrated by 
the empirical binding energies for a given NN interaction potential.  
Then, the evaluation of further experimental observables of nuclei such us radii and also exotic systems such us hypernuclei could be attempted with this type of procedure~\cite{aka2}.

In the present approach, nucleons are treated as classical, 
structureless particles.  
The model Hamiltonian consists of nucleon kinetic 
energy, NN ($V_{NN}$), 
Coulomb ($V_{Coul}$) and Pauli ($V_{Pauli}$) potentials.
The Pauli potential simulates nucleon fermionic nature    
using the Gaussian form introduced by Dorso {\it et al.}~\cite{dorso}, 
but we allow for a density dependence.
In this study, we use a simplified NN interaction potential keeping only 
S-wave interactions without isospin nor spin dependence~\cite{aka}. 
The model Hamiltonian is given by, 
\begin{equation}
H=\sum_{i=1}^{A} \frac{{\bf p}_{i}^{2}}{2m_N}
+ \sum_{i=1,j>i}^{A} 
\left[ {V}_{NN}(r_{ij})+V_{Coul}(r_{ij})
+V_{Pauli} (r_{ij},p_{ij}) \right], 
\label{SWCP}
\end{equation}
where ${\bf p}_{i}$ is the 3-momentum of $i$-th nucleon and 
$r_{ij}=|{\bf r}_i-{\bf r}_j|$ ($p_{ij}=|{\bf p}_i-{\bf p}_j|$) 
the relative distance (momentum) of the $i$-th and $j$-th nucleons.
Explicit expressions for the potentials in 
Eq.~(\ref{SWCP}) are as follows.

{\it $\bullet$ NN interaction potential}: 
\begin{equation}
V_{NN}(r_{ij})=
\left\{
\begin{array}{lll}
V_{Core}, &\quad {\rm for} \quad 0 \le r_{ij} < a,  
&\quad \\
-V_0,     &\quad {\rm for} \quad a \le r_{ij} < b,
&\quad \\
0,        &\quad {\rm for} \quad a+b \le r_{ij}. &
\end{array}\right. 
\label{NNpot}
\end{equation}
The potential consists of a repulsive core of strength $V_{Core}$  of width 
$a$ and an attractive well of strength $V_0$ and width $b$. 
The values used are, $V_{Core}=10$ MeV, $a=1$ fm and $b=2$ fm. 
For $V_0$, we use two values, $V_0=3$ MeV and $V_0=5$ MeV.

{\it $\bullet$ Coulomb potential}:
\begin{equation}
V_{Coul}(r_{ij})=  
\frac{e^2}{4 \pi r_{ij}} (1/2+\tau_i)(1/2+\tau_j),   
\label{Coulombpot}
\end{equation}
where $\tau_i$ ($\tau_j$) is the isospin third-component of
$i$-th ($j$-th) nucleon ($+1/2$ for protons, $-1/2$ for neutrons),  
and $e$ the proton electric charge. 

{\it $\bullet$ Pauli potential}:
\begin{equation}
{V}_{Pauli}({r}_{ij}, {p}_{ij})= 
V_P\,\,  \exp \left(-\frac{r_{ij}^2}{2q_0^2}
-\frac{p_{ij}^2}{2p_0^2}\right) \delta_{\tau_i \tau_j}
\delta_{\sigma_i \sigma_j}, 
\label{Paulipot}
\end{equation}
where $\delta_{\tau_i \tau_j}$ ($\delta_{\sigma_i \sigma_j}$) is  
the Kronecker's delta for the isospin (spin) third-component. 
It prevents nucleons from occupying the same phase 
space volume when they have the same 
quantum numbers. (See Ref.~\cite{newpauli} for other approaches.)
As demonstrated in Ref.~\cite{aka}, 
it is crucial to allow a density dependence for 
this Pauli potential if one wants 
to reproduce the empirical binding energies.
Thus, we discuss next the density dependence of the Pauli potential 
before presenting results.

\begin{figure}[hbtp]
\begin{center}
\includegraphics [angle=-90,scale=1] {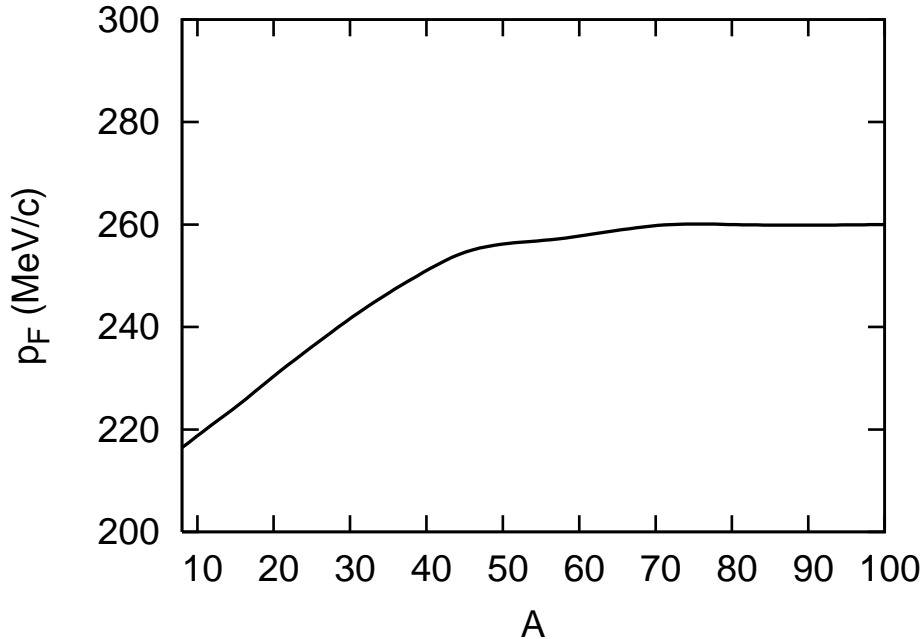}
\caption{Fermi momentum interpolated vs. nucleon number $A$.
}
\label{kf_A}
\end{center}
\end{figure}

First, we show in Fig.~\ref{kf_A} the average Fermi momentum 
versus the nucleon number $A$ obtained by interpolating 
the values given in Ref.~\cite{moniz}.
%%%%%%%%%%%%%%%%%%%%%%%%%%%%%%%%%%%%%%%%%%%%%%%%%%%%%%%%%%%%%%%%%
%%%%%%%%%%%%%%%%%%%%%%%%%%%%%%%%%%%%%%%%%%%%%%%%%%%%%%%%%%%%%%%%%
One can see that the Fermi momenta ($p_F$) increase as the nucleon number 
$A$ increases up to $A \simeq 50$.
For heavier nuclei with $A > 50$ 
they saturate to a value of $p_F \simeq 260$ MeV/c 
(e.g., $265$ MeV/c for a $^{208}$Pb nucleus)~\cite{moniz}.
Thus, including also heavy nuclei with $A > 50$, 
it is more convenient to use the nucleon number $A$ to   
characterize the density dependence of the 
Pauli potential, although the Fermi momentum was used 
previously~\cite{aka} for nuclei with $8 \le A \le 44$.
In this work, we need to extend the parametrization to heavier systems 
to parameterize the density dependent strength 
$V_P$ in the Pauli potential Eq.~(\ref{Paulipot}). 

For $q_0$ and $p_0$ in the Pauli potential, the density 
dependence can be determined as follows.
In a nucleus, a typical nucleon sphere radius $r$ may be given by,
\begin{equation}
r=\left(\frac{3}{4 \pi \rho}\right)^{1/3},
\label{distance}
\end{equation}
where $\rho=2 p_F^3/3 \pi^2$ is the nucleon density and
$p_F$ the nucleon Fermi momentum.
Then, the average inter-nucleon distance $2r$
may be estimated as $(2r/\sqrt{2}q_0)\simeq 1$,
where $q_0$ is "an effective range" of the Pauli potential.
With the uncertainty principle, $q_0\,\, p_0 \simeq \hbar$, 
this leads to:
\bg
q_0&\simeq&\frac{(9 \pi)^{1/3}\hbar}{\sqrt{2}p_F},
\label{q0}\\
p_0&\simeq&\frac{\hbar}{q_0}
=\frac{\sqrt{2}}{(9 \pi)^{1/3}}p_F. 
\label{p0}
\en
For the Fermi momentum $p_F$ appearing in Eqs.~(\ref{q0})
and~(\ref{p0}), we use the value as shown in
Fig.~\ref{kf_A} for the nuclei with $8 \le A \le 44$,
while for $48 \le A \le 100$, we use the saturated value, 
$p_F=260$ MeV/c. 

\begin{figure}[hbtp]
\begin{center}
\includegraphics [angle=-90,scale=1] {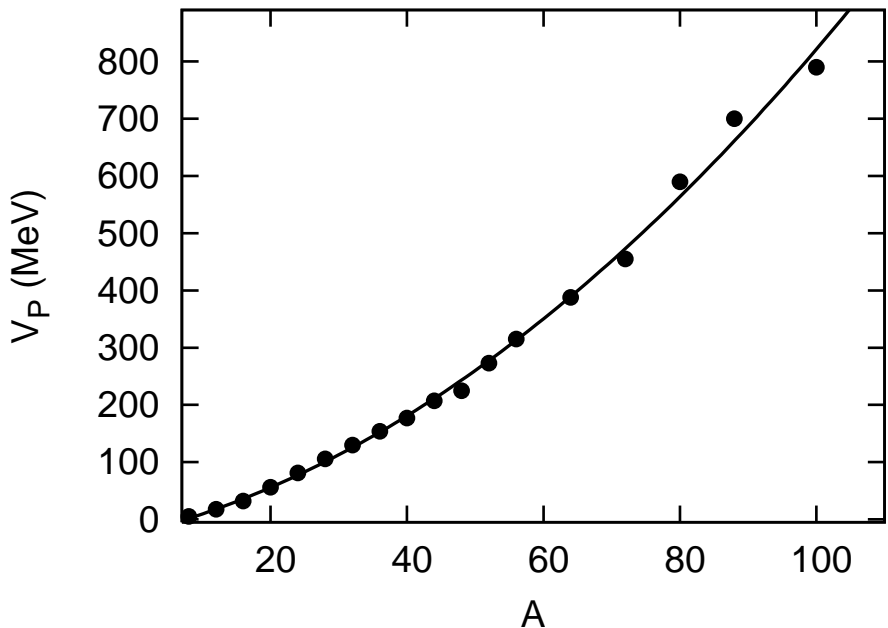}
\caption{Pauli potential strength $V_P$ vs. nucleon number $A$ 
obtained with $V_0=3$ MeV in the NN potential Eq.~(\ref{NNpot}).
Simulations are performed for nuclei with $A=8,12,16, ..., 100$ 
(even $Z$ and $N$) spin-isospin saturated nuclei. 
This applies for all results treated in this study.
The solid line corresponds to the parameterization 
given by Eq.~(\ref{VP_A3}).
}
\label{VP_A}
\end{center}
\end{figure}

By performing simulations to reproduce the empirical binding energies 
for the nuclei with $8 \le A \le 100$ for $V_0=3$ MeV 
in Eq.~(\ref{NNpot}), 
we get the $A$ dependence for $V_P$ in the 
Pauli potential as shown in Fig.~\ref{VP_A} (the blobs).
%%%%%%%%%%%%%%%%%%%%%%%%%%%%%%%%%%%%%%%%%%%%%%%%%%%%%%%%%%%%%%%%%
%%%%%%%%%%%%%%%%%%%%%%%%%%%%%%%%%%%%%%%%%%%%%%%%%%%%%%%%%%%%%%%%%
This justifies that the density dependence of the
Pauli potential may be parameterized well by the nucleon number $A$.  
As expected, the strength $V_P$ of the Pauli potential 
increases with $A$. 
This behaviour may be analogous to
that of the vector potential in Hartree approximation
in relativistic mean field models~\cite{QHD}.
Then, for $V_0=3$ MeV, we get the parameterization: 
\bge
V_P(A)=-25.645+2.9596A+0.0551A^2 \rm\,\ (MeV), 
%V_P(p_F(A))=-25.645+2.9596A+0.0551A^2 \rm\,\ (MeV),
\label{VP_A3}
\ene
\noindent
while for $V_0=5$ MeV, we get:
\bge
V_P(A)=-1088.2+140.55A+0.9809A^2 \rm\,\ (MeV). 
%V_P(p_F(A))=-1288.1969+140.5495A+0.9809A^2 \rm\,\ (MeV), 
\label{VP_A5}
\ene
%
%With these parameterizations, we can model 
%nuclei with $8 \le A \le 100$, corresponding to $V_0=3$ MeV and 
%$V_0=5$ MeV in the NN interaction potential Eq.~(\ref{NNpot}), 
Both cases reproduce the empirical binding energies well. 
Note that the parameterizations are approximate and are given 
as a guidance.
Thus, this suggests that, for a given reasonable  
NN interaction potential, we can model a set of nuclei which 
are calibrated by the empirical binding energies.
Then, we can use them to study other properties of these nuclei,   
such as charge distribution, proton and neutron density distributions, 
and proton and neutron r.m.s. radii~\cite{aka2}.
Although this procedure may be simple, one can regard 
that all complicated many-body effects 
are condensed into a density dependent effective Pauli potential.
The present approach is not based on the fundamental 
theory of strong interaction QCD, but we would like to emphasize, 
a simple, pragmatic aspect for treating complicated, 
many-nucleon systems in semiclassical simulations.

Now, we are in a position to discuss the results. 
In the present study, all simulations are performed 
with a fixed temperature $T=1$ MeV.  
In the simulation a nucleus is constructed 
by initially placing $A$ nucleons uniformly inside 
a sphere of radius $R_0$ of range 2-3 fm 
within a cubic box of volume $V=L^3$ and impose $L >> r_{ij}$.
Then, using the Metropolis algorithm~\cite{Metropolis}, 
the ground state configuration 
is searched by thermal relaxation. 
The Pauli potential should be gradually turned on 
to avoid instabilities.
After this, we sample the configurations in order to calculate 
the statistical average for its binding energy.

\begin{figure}[hbtp]
\begin{center}
\includegraphics [angle=0,scale=2] {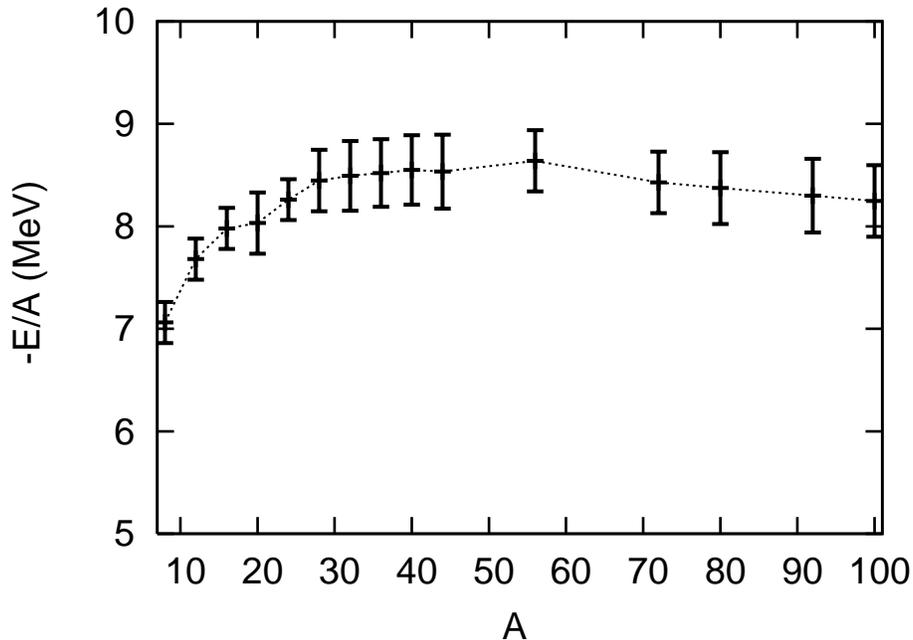}
\caption{Binding energy per nucleon, $\be$, for the set of nuclei considered in this work. (See also caption of Fig.~\ref{VP_A}.)
}
\label{BE}
\end{center}
\end{figure}

\begin{figure}[hbtp]
\begin{center}
\includegraphics [angle=-90,scale=1] {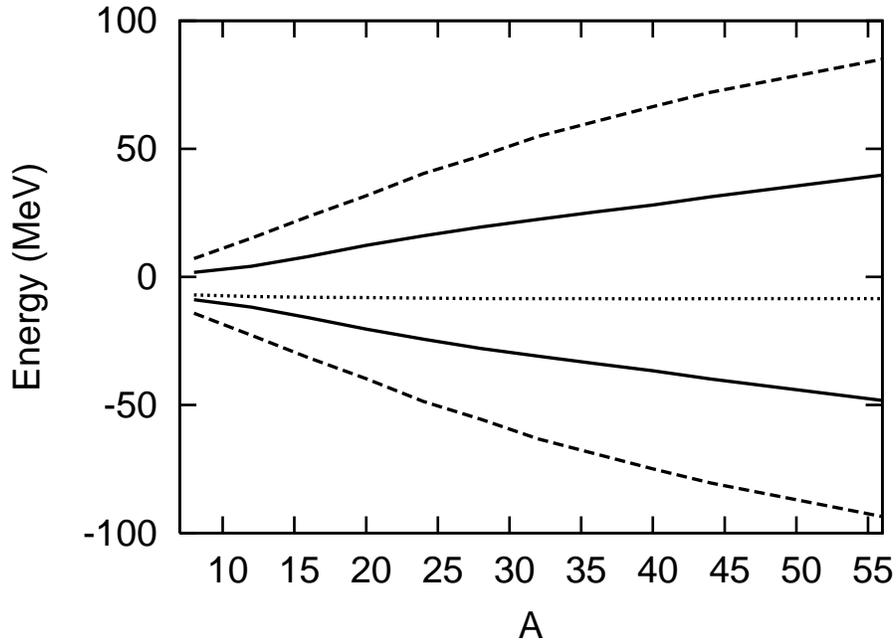}
\caption{Kinetic and potential energy contributions to the binding energy 
per nucleon. The solid and dashed lines are the results 
obtained with $V_0=3$ and $V_0=5$ MeV, respectively. 
For each case, the upper (lower) line corresponds to the contribution from 
the kinetic (potential) energy per nucleon. 
The dotted line is the sum of the two contributions 
for both cases, and goes trough the empirical values.
(See also caption of Fig.~\ref{VP_A}.)
}
\label{scale}
\end{center}
\end{figure}

In Fig.~\ref{BE} we show the binding energy per nucleon 
($\be$) as calculated with the model with $V_0=3$ MeV versus nucleon number $A$. The central bars show the calculated values of nuclei considered in this work and match those in Ref.~\cite{nucdata}. The dotted line is a guidance for eyes.
%%%%%%%%%%%%%%%%%%%%%%%%%%%%%%%%%%%%%%%%%%%%%%%%%%%%%%%%%%%%%%%%%
%%%%%%%%%%%%%%%%%%%%%%%%%%%%%%%%%%%%%%%%%%%%%%%%%%%%%%%%%%%%%%%%%
The statistical uncertainties are shown by error-bars, and 
they are less than 5 \%.
The empirical values are well reproduced with the 
density dependent Pauli potential.  
Thus, for this NN interaction potential with 
$V_0=3$ MeV in Eq.~(\ref{NNpot}), we have obtained a set of nuclei 
which reproduce the empirical binding energies. Let us note that shell effects are partially included through the $A$ dependence in the Pauli potential strength parameterization. 
Next, we discuss the dependence on the NN interaction potentials 
by comparing the results with $V_0=3$ MeV and $V_0=5$ MeV. 
We show contributions from the kinetic and potential energies 
in Fig.~\ref{scale} for the nuclei with $8 \le A \le 56$.
%%%%%%%%%%%%%%%%%%%%%%%%%%%%%%%%%%%%%%%%%%%%%%%%%%%%%%%%%%%%%%%%%
%%%%%%%%%%%%%%%%%%%%%%%%%%%%%%%%%%%%%%%%%%%%%%%%%%%%%%%%%%%%%%%%%
The solid and dashed lines are the results
for $V_0=3$ and $V_0=5$ MeV, respectively.
For each case, the upper (lower) line corresponds the contribution from
the kinetic (potential) energy per nucleon.
The dotted-line is the sum of the two contributions, 
for {both} $V_0=3$ MeV and $V_0=5$ MeV cases.
This suggests that the empirical binding energies 
can be always reproduced by introducing a proper counter balancing Pauli 
potential, once a NN interaction is specified.

\begin{figure}[hbtp]
\begin{center}
\includegraphics [angle=-90,scale=1] {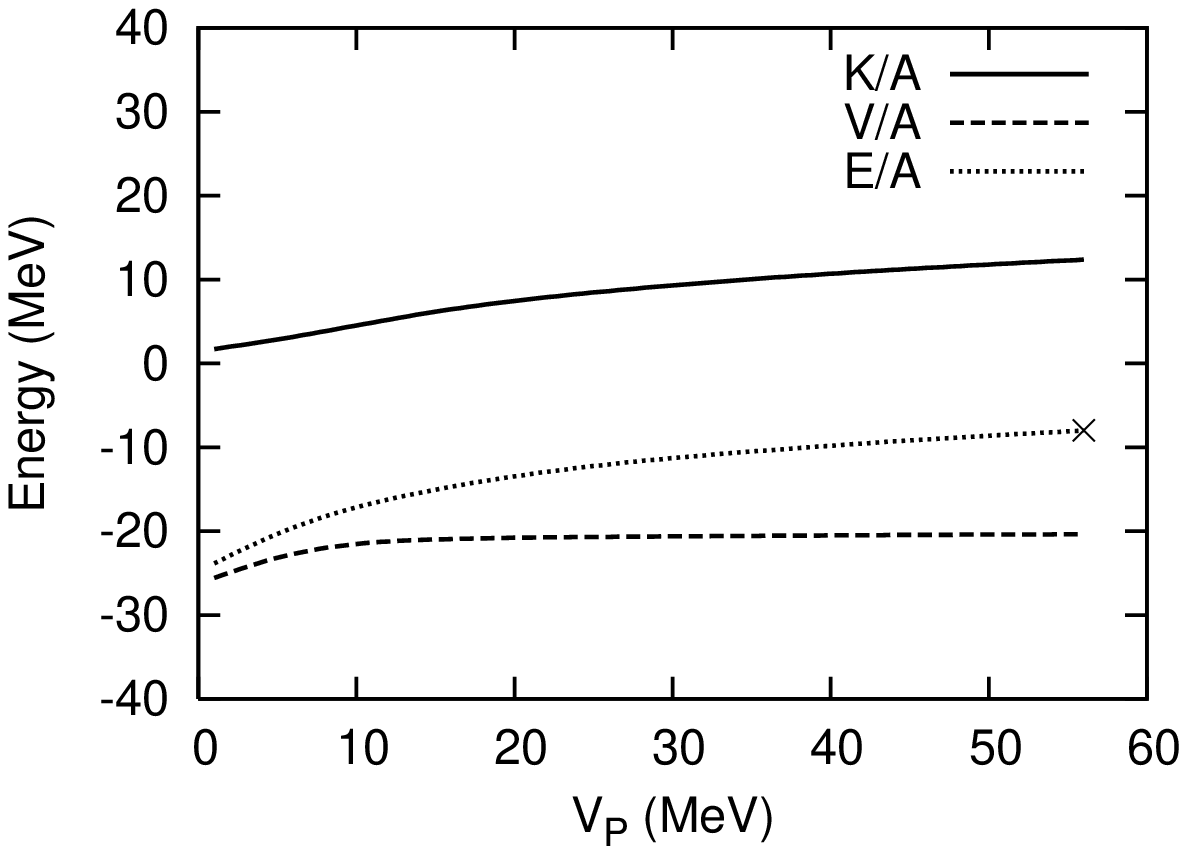}
\caption{Energy contribution to the $^{20}$Ne total energy.}
The solid (dashed) [dotted] line
stands for the kinetic (potential) [total] energy per
nucleon, and labelled by $K/A$ ($V/A$) [$E/A$].
The empirical value is indicated by the cross.
\label{ENe20}
\end{center}
\end{figure}

Further, we analyze how the empirical 
binding energy can be achieved by considering the $^{20}$Ne 
nucleus case as an example. In Fig.~\ref{ENe20} we show kinetic ($K/A$), 
potential ($V/A$) and total ($E/A$) energies per nucleon 
for a $^{20}$Ne nucleus versus the Pauli potential 
strength $V_P$.
%%%%%%%%%%%%%%%%%%%%%%%%%%%%%%%%%%%%%%%%%%%%%%%%%%%%%%%%%%%%%%%%%
%%%%%%%%%%%%%%%%%%%%%%%%%%%%%%%%%%%%%%%%%%%%%%%%%%%%%%%%%%%%%%%%
The empirical value is indicated by the cross.
As increasing the Pauli potential strength $V_P$, 
the kinetic energy contribution increases, 
while the potential energy contribution stays nearly constant.
The increase of the kinetic energy contribution originates 
from the momentum dependence in the Pauli potential, 
due to the modification in the canonical momenta (or the effective masses) 
of the interacting nucleons via the Pauli potential. 
Because of this positively increasing kinetic 
energy, the empirical binding 
energy is finally achieved.
This example is for a fixed nucleon number nucleus.
In order to be able to reproduce the empirical binding energies 
for a set of nuclei, the similar procedure must be repeated 
for all the nuclei in the set.
Thus, one can naturally understand why $A$ (or density)  
dependence in the Pauli potential is necessary.

For the first time we have demostrated that the density dependence of the Pauli potential, crucial to reproduce the empirical binding energies in semiclassical simulations, can be parametrized in terms of a single variable: the nucleon number A. Such results overcomes previous hipothesis describing the density dependence in terms of the Fermi momentum. Once the correct density dependence is included through the Pauli potential, the only other necessary ingredient is a reasonable nucleon-nucleon interaction. The procedure presented is robust, the simulation remaining stable when incresing the number of particles. The validity of the parametrization of the Pauli potential in terms of the nucleon number A is made manifest by the counterbalance between the growing repulsion coming from the kinetic energy and the incresing attraction of the potential energy. This result opens the door to study properties of wider sets of nuclei with a correct parametrization of the Pauli potential. 
Although the procedure presented in this work is 
not based on the fundamental theory of strong interaction 
QCD, we would like to emphasize that, it 
suggests a simple, pragmatic procedure in modelling 
a set of nuclei calibrated by the empirical binding energies 
for a given NN interaction potential. 
Then, using this procedure one may think of further testing asymmetric systems
by studying other properties of nuclei such as radii or other exotic nuclei
in semiclassical simulations.

%%%%%%%%%%%%%%%%%%%%%%%%%%%%%%%%%%%%%%%%%%%%%%%%%%%%%%%%%%%%%%%%
\vspace{2ex}
\noindent{\bf Acknowledgments}\\
We would like to thank Tomoyuki Maruyama and A.W. Thomas for 
helpful discussions. M.A.P.G. would like to dedicate this work 
to the memory of J.M.L.G. This work has been partially funded by 
the Spanish Ministry of Education and Science projects 
DGI-FIS2006-05319, SAB2005-0059 and FPA2004-05616,
and by Junta de Castilla y Le\'on under contracts SA-104104, 
SA-104/04, and EPA-2004-05616.
%%%%%%%%%%%%%%%%%%%%%%%%%%%%%%%%%%%%%%%%%%%%%%%%%%%%%%%%%%%%%%%%
% Bibliography. 
%%%%%%%%%%%%%%%%%%%%%%%%%%%%%%%%%%%%%%%%%%%%%%%%%%%%%%%%%%%%%%%%

%%%%%%%%%%%%%%%%%%%%%%%%%%%%%%%%%%%%%%%%%%%%%%%%%%%%%%%%%%%%%%%%
\end{document}